\documentclass[aps,pre,reprint,superscriptaddress,amsmath,amssymb,showpacs,floatfix,twocolumn]{revtex4-2}
\usepackage{graphicx}
\usepackage{bm}
\usepackage{color}
\usepackage{soul}
\usepackage{dcolumn}
\usepackage{multirow}
\usepackage{tikz}
\usepackage{hhline}
\usepackage{mathtools}
\usepackage{algorithm}
\usepackage{algorithmic}
\usepackage[normalem]{ulem}
\usepackage[colorlinks, linkcolor=blue, anchorcolor=blue, citecolor=blue, urlcolor=blue]{hyperref}

\newcommand{\tr}[1]{{\rm tr} \{#1\} }
\newcommand{\ket}[1]{|#1\rangle}
\newcommand{\bra}[1]{\langle#1|}

\begin{document}
\title{Estimating Entanglement Entropy via Variational Quantum Circuits with Classical Neural Networks}
\author{Sangyun Lee}
\affiliation{Institute for Physical Science and Technology, University of Maryland, College Park, Maryland 20742, USA}
\affiliation{School of Physics, Korea Institute for Advanced Study, Seoul, 02455, Korea}
\author{Hyukjoon Kwon}
\email{hjkwon@kias.re.kr}
\affiliation{School of Computational Sciences, Korea Institute for Advanced Study, Seoul 02455, Korea} 
\author{Jae Sung Lee}
\email[]{jslee@kias.re.kr}
\affiliation{School of Physics, Korea Institute for Advanced Study, Seoul, 02455, Korea}

\begin{abstract}
Entropy plays a crucial role in both physics and information science, encompassing classical and quantum domains. In this work, we present the Quantum Neural Entropy Estimator (QNEE), a novel approach that combines classical neural network (NN) with variational quantum circuits to estimate the von Neumann and R\'enyi entropies of a quantum state. QNEE provides accurate estimates of entropy while also yielding the eigenvalues and eigenstates of the input density matrix. Leveraging the capabilities of classical NN, QNEE can classify different phases of quantum systems that accompany the changes of entanglement entropy. Our numerical simulation demonstrates the effectiveness of QNEE by applying it to the 1D XXZ Heisenberg model. In particular, QNEE exhibits high sensitivity in estimating entanglement entropy near the phase transition point. We expect that QNEE will serve as a valuable tool for quantum entropy estimation and phase classification.
\end{abstract}
\maketitle

\section{introduction}
Entropy, originally introduced in classical thermodynamics and statistical physics, has evolved into a fundamental concept used to quantify information across various scientific disciplines, ranging from computer science to classical and quantum information theory. In particular, the Shannon entropy establishes a profound connection between thermodynamics and information theory, which is further extended to quantum systems through the introduction of the von Neumann entropy. This extension has paved the way for the exploration of intricate relationships between information and physical reality, as demonstrated by the discovery of information fluctuation theorems~\cite{Sagawa2010,Sagawa2022} and Landauer's principle~\cite{Landuer, LandauerNature, Proesmans, LeePRL2022} in both classical and quantum domains. Moreover, in the realm of quantum information theory, entropy has been employed to unveil non-classical characteristics inherent to quantum systems. Specifically, the entanglement entropy captures non-classical correlations between distinct parties of a quantum system and plays a pivotal role in distinguishing different phases of quantum materials~\cite{Osterloh2002Nat, vidal2003entanglement, Eisert2010RMP, Osborne2022PRA,
singh2023inferring}.

In spite of its recognized importance as a valuable theoretical tool, the practical determination of entropy values in physical systems presents substantial challenges especially for a high-dimensional system. This challenge is similarly encountered when estimating stochastic entropy production, as it is determined by the probability density of the system's trajectory~\cite{seifert2012stochastic, Roldan2010Estimating}. 
Physicists have addressed this difficulty through  entropy-production inequalities, such as the thermodynamic uncertainty relation and entropic bound~\cite{Fakhri2019, otsubo2022estimating, SYLee2023}. 
In the realm of quantum systems, estimating entropy becomes even more formidable due to the dependence on measurement choices. Specifically, the von Neumann entropy of a quantum state is determined as the minimum entropy among all potential measurement bases, which is achieved by taking the state's eigenbasis. This additional layer of complexity introduces difficulties in obtaining the eigendecomposition of a quantum density matrix, with computational costs exponentially increasing with the size of the quantum system. While several indirect methods have been proposed~\cite{vanEnk2012, Abanin2012, Elben2018, Kwon2020} and experimentally demonstrated~\cite{Islam2015, Adam2016, Linke2018, Brydes2019} to estimate or bound entanglement of quantum states, limitations in accuracy persist. Recently, variational quantum algorithms (VQA) \cite{Cerezo2021VQA}, such as the variational quantum state diagonalization (VQSD)~\cite{larose2019variational} and the variational quantum state eigensolver (VQSE)~\cite{cerezo2022variational} have emerged as potential approaches for estimating quantum state eigenvalues. However, it still remains uncertain whether these methods can efficiently estimate quantum entropy.

The challenging nature associated with quantum entropy estimation prompts the pursuit of a novel and efficient algorithm that can be effectively integrated with a variational quantum circuit, leading to improved performance. To this end, one potential candidate is classical neural networks (NNs), which have demonstrated exceptional capabilities in optimizing objective functions. Notably, this machine-learning technique has recently found application in estimating entropy production for classical thermodynamic systems, named as neural estimator of entropy production (NEEP) algorithm~\cite{kim2020learning, otsubo2022estimating}, by optimizing the objective function defined by the Donsker-Varadhan inequality~\cite{Donsker}. This utilization of NNs in classical thermodynamics provides inspiration for their potential application in the context of quantum entropy estimation. By leveraging the strengths of classical NNs in optimizing von Neumann entropy, it would be possible to enhance the accuracy and efficiency of quantum entropy estimation, paving the way for further advancements in the field.

In this work, we present a quantum-classical hybrid algorithm, Quantum Neural Entropy Estimator (QNEE), for estimating the von Neumann and R\'enyi entropies of a quantum state. Our method capitalizes on the efficient entropy estimation capabilities of NNs from the outcomes of parametrized quantum circuits while utilizing variational parameters to align the quantum circuit in the correct measurement basis. We highlight that our methodology can be readily extended for any quantum circuit ansatz as the classical NN is constructed independent of the variational quantum circuit's structure. 

Classical NN using the cost function based on the Donsker-Varadhan inequality has some useful properties, such as robustness against small-size data and rare events, which have already been verified in classical stochastic processes~\cite{kim2020learning,nir2020machine, SYLee2023}. Another advantage of this type of cost function is that it can provide a reliable upper bound of the entropy even if the measurement basis is not fully aligned correctly. This opens a possible application of the proposed approach as a phase classifier of quantum systems even without the accurate estimation of entanglement value.

To showcase the effectiveness of our proposed protocol, we apply it to estimate the ground state entanglement entropy of the Heisenberg XXZ model and to distinguish its phase relying on the external magnetic field. Our numerical results demonstrate that QNEE exhibits good performance in estimating entanglement entropy as well as the subsystem's eigenvalues. We also provide a detailed comparison between QNEE and the existing scheme of VQSE.

This paper is organized as follows. In Sec.~\ref{sec:QNEE_protocol}, we construct QNEE by combining a classical NN with a variational quantum circuit. In Sec.~\ref{sec:result}, we apply QNEE to estimate the ground state entanglement entropy of the 1D XXZ Heisenberg model and its performance as a phase classifier. We conclude the paper in Sec.~\ref{sec:disscussion}.

\vspace{0.25cm}
\textbf{Concurrent work:} In the course of finalizing this manuscript, we became aware of related research that has emerged in arXiv~\cite{shin2023estimating, goldfeld2023quantum}. It is essential to highlight that our work draws inspiration from the authors' earlier contributions~\cite{kim2020learning, kim2022estimating, SYLee2023} as its quantum extension. Therefore, our cost function for NN is distinct from that employed in the concurrent works. Furthermore, we must emphasize that the primary focus of our work lies in the application of phase classification to a physical Hamiltonian. In contrast, the concurrent works under examination delve into the investigation of random quantum states as illustrative examples. These distinctions in scope and approach further underscore the unique contributions and significance of our study.

\section{quantum entropy estimator}
\label{sec:QNEE_protocol}

\subsection{Cost function of QNEE}
In this section, we establish QNEE's cost function to estimate the quantum entropy of a given density matrix $\hat \rho$, describing the state of a system. The cost function is based on the following Gibbs variational principle:
\begin{align}
    S(\hat \rho ) \equiv - {\rm tr} \{\hat \rho \ln{\hat \rho}\} \leq -\langle \hat O\rangle_{\hat \rho} +\ln{ {\rm tr} \{e^{\hat O}\}  }
\label{eq:GibbsVar_main}\end{align} 
for any Hermitian operator $\hat O$, whose expectation value is defined as $\langle \hat O \rangle_{\hat \rho} = {\rm tr}\{  \hat O\hat \rho\}$. We present a simple derivation of Eq.~\eqref{eq:GibbsVar_main} in appendix~\ref{sec:derivation}. Equation~\eqref{eq:GibbsVar_main} was utilized as a quantum entropy estimator in Ref.~\cite{goldfeld2023quantum}. However, as explained in Sec.~\ref{sec:opt_classical} and appendix~\ref{sec:derivation}, a cost function, including the logarithmic function, can be biased and malfunction for large data if additional manipulation is not employed.
To avoid this complication, we further linearize Eq.~\eqref{eq:GibbsVar_main} using the inequality $\ln {\rm tr} \{e^{\hat O}\} \leq {\rm tr} \{e^{\hat O}\}-1 $ as
\begin{align}
    S(\hat \rho ) \leq -\langle \hat O\rangle_{\hat \rho} + {\rm tr} \{e^{\hat O}\} - 1. 
\label{eq:costfunction}\end{align}
Equation~\eqref{eq:costfunction} is usually a more appropriate cost function for the stochastic (mini-batch) gradient method~\cite{ruder2016overview}, which is typically used for large-system NN optimization~\cite{belghazi2018mine}.
The both bounds~\eqref{eq:GibbsVar_main} and 
\eqref{eq:costfunction} imply that we can indirectly estimate the von Neumann entropy using the operator $\hat O$'s expectation value and its eigenvalues. However, the operator saturating the both bounds in Eqs.~\eqref{eq:GibbsVar_main} and \eqref{eq:costfunction} is given by $\hat O^* = \ln{\hat \rho}$. In other words, the exact estimation using these bounds requires the information of $\hat \rho$. Nevertheless, an optimization method enables us to estimate the von Neumann entropy without accessing full information of the density matrix $\hat \rho$. For classical stochastic processes, various optimization methods have been employed to estimate the entropy production without detailed information~\cite{manikandan2019inferring,kim2020learning,otsubo2020estimating,SYLee2023,Kwon2023Divergence}. These series of studies have shown that NN-based approaches yield reliable and robust estimation against small-size data or data with rare event.

We employ a parametrized quantum circuit and a classical NN to find the optimal operator $\hat O$ for an $n$-qubit quantum state $\hat \rho$. To this end, we rewrite Eq.~\eqref{eq:costfunction} using the matrix diagonalization of the observable~$\hat O= \hat V^\dagger  \hat D \hat V $ with the diagonal matrix~$\hat D$ and the unitary matrix~$\hat V$, which leads to
\begin{align}
    S(\hat \rho) \leq& - \langle \hat D \rangle_{\hat{\rho}_{\hat V}} + {\rm tr} \{ e^{\hat D} \} - 1,
\end{align}
where $\hat \rho_{\hat V} = \hat V \hat \rho \hat V^\dagger$. The optimization over $\hat O$ then becomes finding the optimal diagonal matrix $\hat D$ and the unitary operator $\hat V$. By taking $\hat D(\Theta_N) = \sum_{i=1}^{2^n} h(s_i;\Theta_N) \ket{s_i}\bra{s_i}$ in the computational basis for $n$ qubits $\{ \ket{s_i} \} = \{ \ket{00\cdots 0}, \cdots, \ket{11\cdots 1} \}$, we encode $h(s_i;\Theta_N)$ into the output function of a classical NN with a weight set  $\Theta_N$. We also encode the unitary operator $\hat V(\Theta_Q)$ into a variational quantum circuit parameterized by a vector $\Theta_Q$. The QNEE's cost function $C(\Theta_N, \Theta_Q)$ is then defined as 
\begin{align}
    S(\hat \rho) \leq&  -\sum_{i=1}^{2^n} h(s_i;\Theta_N) P_{\hat V}(s_i;\Theta_Q) + \sum_{i=1}^{2^n}e^{h(s_i;\Theta_N)} -1 \nonumber\\
    \equiv& C(\Theta_N, \Theta_Q),
\label{eq:VNcostftn}
\end{align} 
where $ P_{\hat V}(s_i;\Theta_Q) = \langle s_i | \hat V(\Theta_Q) \hat \rho \hat V^\dagger (\Theta_Q) |s_i \rangle $ is the probability of the quantum circuit parameterized by $\Theta_Q$ to output the string $s_i$. Here we define $\hat V_{\rm D}$ as the unitary matrix rendering $\hat \rho$ diagonalized as $\rho_{\hat V_{\rm D}} = \hat V_{\rm D} \hat \rho \hat V_{\rm D}^\dagger = \sum_i \lambda_i |s_i\rangle \langle s_i|$, where $\lambda_i = \langle s_i|\hat \rho_{\hat V_{\rm D}}| s_i\rangle$ is the probability for measuring $s_i$ from $\hat \rho_{\hat V_{\rm D}}$. When $\hat V (\Theta_Q) = \hat V_{\rm D}$ and $h(s_i;\Theta_N) = \ln \lambda_i$, the inequality in Eq.~\eqref{eq:VNcostftn} is saturated, in which case $P_{\hat V} (s_i;\Theta_Q) = \lambda_i$.

Equation~\eqref{eq:VNcostftn} is reminiscent of the cost function of NEEP: a neural estimator for entropy production~\cite{kim2020learning}. Both the cost functions of QNEE and NEEP can be derived from the $f$-divergence~\cite{Basu1998Robust, Kwon2023Divergence}. 
The unsupervised learning with a variational cost function has been established as an effective method to estimate the divergence of data rather than other methods, especially when the number of samples is inadequate or the data include rare events~\cite{kim2020learning, SYLee2023}. Because the entropy can be considered as the divergence that one distribution is given as a maximally mixed state, we expect that QNEE would be effective when the given density matrix has a few dominant eigenvalues while the remainings are nearly zero, 
or the number of given shots is insufficient, which is the usual situation for many-qubit systems. The von Neumann entropy is small in general under these conditions. 

The $f$-divergence yields a diverse set of information measures and their variational functional forms. Consequently, this enables the formulation of cost functions for estimating various physical quantities. 
One prominent example of such cost functions is the R\'enyi entropy, defined as
\begin{align}
     S_\alpha (\hat \rho ) = \frac{1}{1-\alpha } \ln{ {\rm tr}\{ \hat\rho^\alpha \} }
    \label{eq:Renyientropy}
\end{align} 
for $\alpha >0$ and $\alpha \neq 1$. The corresponding cost function of Eq.~\eqref{eq:Renyientropy} is
\begin{align}
    \frac{e^{ (1-\alpha) S_\alpha( \hat \rho )  } -1 }{\alpha (1-\alpha)} 
     \leq& 
    \sum_{i=1}^{2^n}
       P_{\hat V}(s_i;\Theta_Q) \frac{ e^{(\alpha -1)h(s_i;\Theta_N) } -1}{(1 - \alpha )}\nonumber\\
    &+\sum_{i=1}^{2^n}
        \frac{e^{\alpha h(s_i;\Theta_N)} - 1}{\alpha}\nonumber\\ 
    \equiv& C_\alpha(\Theta_N, \Theta_Q)
    \label{eq:Renyicost_maintext}
.\end{align}
The detailed derivation of Eq.~\eqref{eq:Renyicost_maintext} from the $f$-divergence is presented in appendix~\ref{sec:Renyiderivation}. It is also straightforward to show that Eq.~\eqref{eq:Renyicost_maintext} is saturated under the same conditions as Eq.~\eqref{eq:VNcostftn}, that is, $h(s_i;\Theta_N) \rightarrow \ln \lambda_i$ and $P_{\hat V} (s_i;\Theta_Q) \rightarrow \lambda_i$.
With the optimal parameters saturating the equality, the cost function can be easily converted into the R\'enyi entropy. Note that the cost function for R\'enyi entropy used in Ref.~\cite{goldfeld2023quantum} consists of the two logarithmic terms, which has the same complication as the Gibbs variational principle~\eqref{eq:GibbsVar_main}, not straightforwardly applicable to the stochastic gradient method to handle large data without additional data manipulation.  
These two logarithmic terms of the cost function in  Ref.~\cite{goldfeld2023quantum} cannot be simply linearized as done in Eq.~\eqref{eq:costfunction} either since it is not trivial to determine which one is larger between the original one and the linearized one. Therefore, it is not easy to find the variational form appropriate for the stochastic gradient method from the cost function in Ref.~\cite{goldfeld2023quantum}. In contrast, our cost function has no such problem, thus, it would be more appropriate for manipulating larger quantum system via NN.

Equation~\eqref{eq:Renyicost_maintext} is reduced to the bound for the von Neumann entropy~\eqref{eq:VNcostftn} in the limit of $\alpha \rightarrow 1$. Moreover,  Eq.~\eqref{eq:Renyicost_maintext}, originally derived for estimating the R\'enyi entropy, can also be applied to estimate the von Neumman entropy even for general $\alpha(\neq 1)$ cases. This is due to the fact that the NN output with optimal parameters $\Theta_N^*$ for the R\'enyi entropy is the same as that of the von Neumann entropy~\eqref{eq:VNcostftn}, i.e.,  $h(s_i;\Theta_N^*)=\ln{\lambda_i}$ for any $\alpha$. This freedom for choosing $\alpha$ arbitrarily opens another possibility to render the NN performance enhanced. A similar technique was used to estimate stochastic entropy production more accurately by tuning $\alpha$~\cite{Kwon2023Divergence}.

\subsection{Overview of the QNEE algorithm}
\begin{figure*}[t]
\centering
\includegraphics[width=0.9\textwidth]{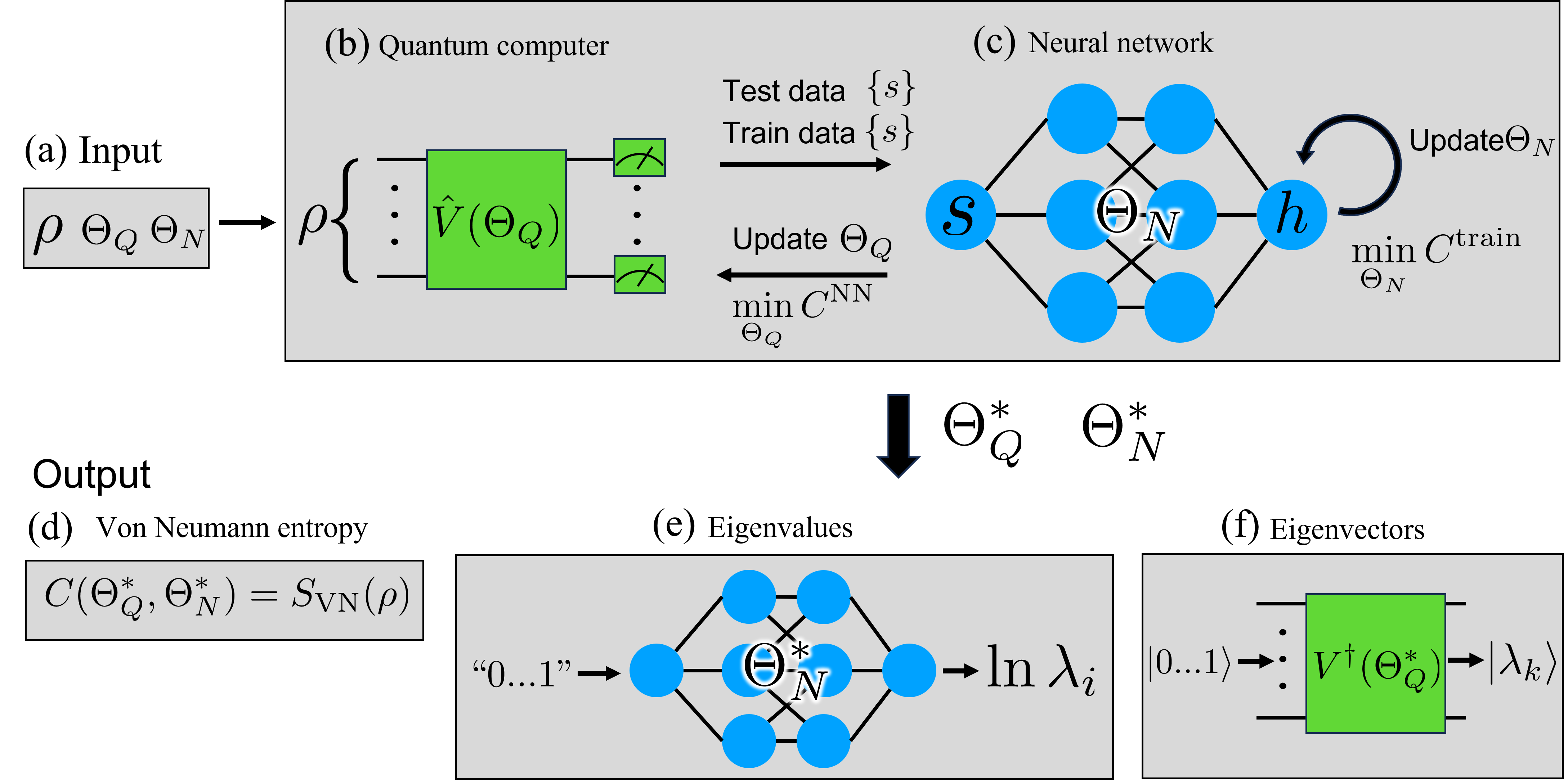}
\caption{ Architecture of QNEE. (a) QNEE requires three inputs: density matrix~$\hat \rho$, initial parameter~$\Theta_Q$ for a quantum circuit, and $\Theta_N$ for a NN. (b) From the given $\Theta_Q$ and density matrix~$\hat \rho$, two string data sets are generated: One for training and the other one for testing. 
(c) We update $\Theta_N$ by minimizing the cost function evaluated with the train data, $\min_{\Theta_N} C^{\rm train}$. During the learning process, the cost function evaluated with the test data~$C^{\rm test}$ is recorded. 
Then, the NN outputs the lowest value of $C^{\rm test}$ from the recorded values, denoted as $C^{\rm NN}$.
The quantum circuit's parameter~$\Theta_Q$ is updated by minimizing the NN's output~$\min_{\Theta_Q} C^{\rm NN}$. 
On successful training, QNEE finds the optimal parameters $\Theta_N^*$ and $\Theta_Q^*$, and QNEE generates three outputs. 
(d) The cost function itself becomes von Neumann entropy with optimal parameters. (e) The trained NN can generate the eigenvalues of $\hat\rho$. (f) With a conjugated quantum circuit~$\hat V^\dagger(\Theta^*_Q)$, eigenvectors can be generated. 
}
\label{fig:concept}
\end{figure*}
Here we present an outline of QNEE's estimation procedure. Figure~\ref{fig:concept} shows the schematic diagram of the comprehensive QNEE algorithm. As depicted in Fig.~\ref{fig:concept}(a), we prepare an $n$-qubit quantum system described by the density matrix $\hat \rho$, for which the von Neumann entropy needs to be estimated. Additionally, we randomly select the initial values of  $\Theta_N$ and $\Theta_Q$, which serve as the parameters for the NN and the quantum circuit, respectively. This setup allows us to commence the repetitive procedure for updating $\Theta_N$ and $\Theta_Q$ until the cost function in Eq.~\eqref{eq:VNcostftn} is optimized. The one unit, as illustrated in Figs.~\ref{fig:concept}(b) and (c), of this repetitive procedure can be divided into two processes in terms of the updating parameters: (i) the updating process of $\Theta_N$ via the optimization of the cost function using the NN and (ii) the updating process of $\Theta_Q$ obtained through the calculation of the gradient of the cost function.

The brief overview of this one-unit process is as follows. First, training and test data sets, consisting of $N_s$ strings each, are produced from the quantum circuit with a given $\Theta_Q$. With the training set, we perform unsupervised learning to update $\Theta_N$ using the classical NN, along with minimizing the cost function. In the course of the $\Theta_N$ updating process, we evaluate the cost function by inputting the test data set into the NN several times and finding the lowest one among them. 
This lowest value of the cost function denoted by $C^{\rm NN}$ and the corresponding parameters $\Theta_N$ are recorded. Details on this NN optimization process and the structure of the NN are presented in Sec.~\ref{sec:opt_classical}. For updating $\Theta_Q$, we need to evaluate the gradient of the cost function with respect to $\Theta_Q$, i.e., $\nabla_{\Theta_Q} C^{\rm NN} $. To do this, it is necessary to obtain $C^{\rm NN}$ for small perturbation $\delta$ from $\Theta_Q$ in various directions through the NN optimization method mentioned above. Employing this gradient descent method, $\Theta_Q$ is updated slightly toward the direction that minimizes $C^{\rm NN}$. Details on the structure of the quantum circuit and the $\Theta_Q$ updating process are explained in Sec.~\ref{sec:opt_quantum}.

Upon conducting an adequate number of repetitions of this unit process to update $\Theta_N$ and $\Theta_Q$, we anticipate the optimization of parameters and cost functions. Utilizing the obtained optimal parameters $\Theta_N^*$ and $\Theta_Q^*$, QNEE produces three essential outputs: the von Neumann entropy ($S(\hat\rho)$), the eigenvalues of the quantum state ($\lambda_i$), and their corresponding eigenvectors ($\ket{\lambda_i}$) as follows:
\begin{align} 
    S(\hat \rho) &=C(\Theta^*_N,\Theta^*_Q),\label{eq:optVonEnt} \\
    \lambda_i &= \exp{[h(s_i;\Theta^*_Q)]}, \label{eq:optEigenvalue} \\
    |\lambda_i \rangle &= \hat V^\dagger(\Theta^*_Q)|s_i\rangle= \hat V^\dagger_D|s_i\rangle. \label{eq:optEigenvector}
\end{align}
We note that the same process can be employed to estimate the R\'enyi entropy with the cost function in Eq.~\eqref{eq:Renyicost_maintext}. 

\begin{figure*}[t]
\centering
\includegraphics[width=0.9\textwidth]{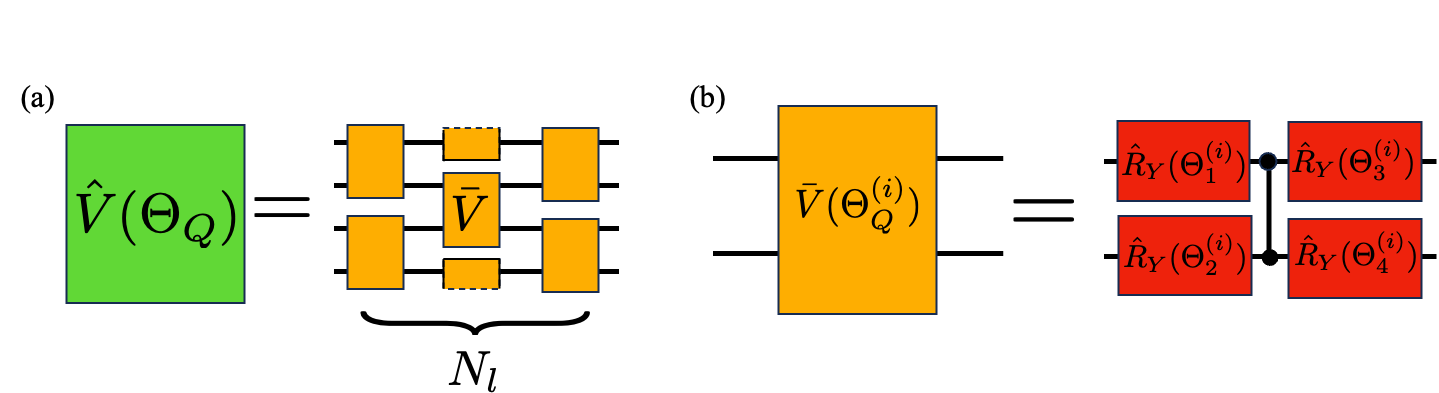}
\caption{ 
Schematic diagrams of (a) layered hardware efficient ansatzs~$\hat V(\Theta_Q)$ and (b) its block gate~$\bar V$. $\hat V(\Theta_Q)$ consists of $N_l$ layers of $\bar V$ block gates. In this research, we choose $\bar V$ as the combination of four $R_Y$ gates and one controlled-Z gate. For $n$ qubits, the dimension of $\Theta_Q$ or the total number of parameters is given as $2 n N_l $.
}
\label{fig:ansatz}
\end{figure*}

\subsection{Structure of the classical NN and optimization of $\Theta_N$}
\label{sec:opt_classical}
In this section, we elaborate the structure and training process of the classical NN. We design the NN as a combination of an embedding layer and fully connected three hidden layers parameterized by $\Theta_N$. The input string is converted into a vector via the embedding layer and is subsequently processed through the hidden layers with 256 units each along with ReLU activation functions~\cite{nair2010rectified}. Therefore, the NN can transform the string data generated via the quantum circuit~$\hat V(\Theta_Q)$ into the scalar value~$h( s;\Theta_N)$. To optimize $\Theta_N$ of the classical NN, we first generate training and test sets consisting of $N_s$ strings each from the quantum circuit parameterized by $\Theta_Q$. Using the training set, we evaluate the cost function in Eq.~\eqref{eq:VNcostftn} as
\begin{align} 
    C^{\rm train} (\Theta_N, \Theta_Q) =& - \frac{1}{N_s} \sum_{s \in {\rm train~set}} h(s;\Theta_N) \nonumber\\
    &+ \sum_{i=1}^{2^n}e^{h(s_i;\Theta_N)} -1 .
\label{eq:fullbatchcost_train}\end{align}
Over a predetermined number of iterations $N_{\rm iter}$, $\Theta_N$ is updated using ADAM optimizer~\cite{kingma2014adam} for minimizing the cost function. 

During these training iterations, we evaluate the cost function several times using the test set with the updated $\Theta_N$ as
\begin{align} \label{eq:fullbatchcost_test}
     C^{\rm test} (\Theta_N, \Theta_Q) =& - \frac{1}{N_s} \sum_{s \in {\rm test~set}} h(s;\Theta_N) \nonumber\\
     &+ \sum_{i=1}^{2^n}e^{h(s_i;\Theta_N)} -1 . 
\end{align}
Among these several evaluated $C^{\rm test}$, the lowest one is recorded as $C^{\rm NN}$ and the corresponding $\Theta_N$ is selected for the NN parameters at a given $\Theta_Q$. Note that the learning curves shown in  Figs.~\ref{fig:eigenvalues}(d-f) are the plot of the recorded $C^{\rm NN}$.

When the NN is ideally trained, the optimal parameter for the NN is given as $\bar \Theta_N (\Theta_Q) = arg\,\min_{\Theta_N} C(\Theta_N,\Theta_Q)$. Then, the output of the NN and the cost function are written as 
\begin{align}
    h(s_i;\bar \Theta_N) =& \ln{ P_{\hat V}(s_i;\Theta_Q) }, \label{eq:intermediateh} \\
    C(\bar \Theta_N,\Theta_Q) =& -\sum_{i=1}^{2^n}  P_{\hat V}(s_i;\Theta_Q)\ln{ \{  P_{\hat V}(s_i;\Theta_Q)\} },
\label{eq:intermediateCostFtn}
\end{align} 
respectively. Equation~\eqref{eq:intermediateCostFtn} is the Shannon entropy~$S[ P_{\hat V} (s_i;\Theta_Q)]$ in terms of the diagonal elements of $\hat \rho_{\hat V} $. Therefore, we can write down the following variational principle with the ideal-string distribution $ P_{\hat V}(s_i;\Theta_Q)$ as 
\begin{align}
    S[ P_{\hat V} (s_i;\Theta_Q)] \leq& -\sum_{i=1}^{2^n} h(s_i;\Theta_N) P_{\hat V}(s_i;\Theta_Q) \nonumber\\
    &+ \sum_{i=1}^{2^n}e^{h(s_i;\Theta_N)} -1
\label{eq:intermediate variationalform}.
\end{align} 
This bound~\eqref{eq:intermediate variationalform} can also be derived from the Donsker--Varadhan inequality~$D_{\rm KL}[P\| Q] \geq \langle h(s) \rangle_P -\ln{ \langle e^{h(s)}\rangle_Q}$ which is the variational form of the Kullback--Leibler divergence~$D_{\rm KL}[P \| Q] = \sum_i P(s_i)\ln{[P(s_i)/Q(s_i)]}$ for two distributions $P(s_i)$ and $Q(s_i)$~\cite{Donsker}. One can easily derive Eq.~\eqref{eq:intermediate variationalform} from the Donsker--Varadahan inequality by choosing the probability $Q$ as a uniform distribution and applying the inequality $\ln{x}\leq x-1$ to the logarithm function.
Note that Eq.~\eqref{eq:intermediateh} reveals the role of the last two terms in $C^{\rm train}$; these terms $\sum_{i=1}^{2^n}e^{h(s_i;\Theta_N)} -1$ ensure that $e^h$ represents a normalized probability.

Finally, it is worth noting that although Eq.~\eqref{eq:fullbatchcost_train} is currently formulated as a full-batch summation, it can be adapted to utilize mini-batch evaluation for calculating the cost function. This implies that the stochastic gradient descent method~\cite{ruder2016overview} can be directly applied to the QNEE's cost function without any modifications, presenting an advantage in using the linearization of the logarithmic function, $\ln x\leq x-1$. The stochastic gradient descent method is efficient in handling large-scale data and facilitating optimizer escape from local minima. Given the exponential increase in data size with the number of qubits, the practical necessity of stochastic gradient descent for managing such vast amounts of data becomes evident. Moreover, we can also extend the application of stochastic gradient descent to the cost function for R\'enyi entropy estimation, as given in Eq.~\eqref{eq:Renyicost_maintext}.

\subsection{Structure of the quantum circuit and optimization of $\Theta_Q$}
\label{sec:opt_quantum}
\label{sec:opt_quantum}
With the cost function calculated by the NN, we update the quantum circuit. Here we explain our choice of ansatz and the optimization process of the quantum circuit. Optimizing $\Theta_Q$ can be viewed as a VQA, the accuracy of which heavily depends on the structure of the target quantum state and quantum circuit ansatz. 
In this work, we choose the layered hardware-efficient ansatz~\cite{kandala2017hardware}, which is widely adopted when intricate information of $\hat \rho$ is inaccessible. The ansatz consists of two-qubit gates of one kind and parameterized one-qubit gates, as described in Fig.~\ref{fig:ansatz}.
$\hat V(\Theta_Q)$ is constructed from $N_l$ layered two-qubit gates~$\bar V(\Theta_Q^{(i)})$ that are drawn as yellow patchs. $\bar V(\Theta_Q^{(i)})$ contains four $\hat R_Y$ gates and one controlled-Z gate, which is the same choice as the quantum circuit in Ref.~\cite{cerezo2022variational}. %
Each quantum circuit layer can be an identity gate with certain parameters to guarantee that increasing $N_l$ improves the accuracy of QNEE~\cite{cerezo2022variational}.

In this study, a gradient descent method is chosen to optimize the cost function via the quantum circuit. To this end, it is necessary to estimate the cost function at differently perturbed $\Theta_Q$s for obtaining its gradient. For example, if the ansatz is parameterized by one single scalar, we estimate $C^{\rm NN}(\Theta_Q)$ and  $C^{\rm NN}(\Theta_Q+\delta)$ via the NN, and calculate the gradient $\nabla_{\Theta_Q} C(\Theta_Q) = [C^{\rm NN}(\Theta_Q+\delta)-C^{\rm NN}(\Theta_Q)]/\delta $, where $\delta$ is a small value. 
Then, the parameter is updated as $ \Theta_Q - \eta_Q \nabla_{\Theta_Q} C(\Theta_Q)$ with a learning rate~$\eta_Q$.
Therefore, two training processes of the NN are required to calculate the gradient for a scalar $\Theta_Q$ case. When the dimension of $\Theta_Q$ increases, the number of necessary directions for perturbation to obtain the gradient also increases. Using the calculated gradient and a learning rate that we choose, $\Theta_Q$ is updated. After an adequate number of iterations of this $\Theta_Q$ updating process, we take the lowest value of the cost function as the estimated von Neumann entropy among all $C^{\rm NN}$ values evaluated during the iteration procedure. 

When the quantum circuit is ideally trained, the optimal parameter is found as $\Theta^*_Q = arg \min_{\Theta_Q} C(\bar \Theta_N,\Theta_Q)$ that diagonalizes the input density matrix with the eigenvalue $\lambda_i =  P_{\hat V_D}(s_i;\Theta^*_Q)$. Therefore, the cost function is saturated to the von Neumann entropy, and the output of the NN with $\Theta^*_N = \arg \min_{\Theta_N} C(\Theta_N,\Theta^*_Q)$ is the logarithm of the eigenvalue as shown in Eqs.~\eqref{eq:optVonEnt} and \eqref{eq:optEigenvalue}, respectively. Additionally, the eigenvectors of $\hat\rho$ can be constructed by the conjugated quantum circuit $\hat V^\dagger (\Theta^*_Q)$ as Eq.~\eqref{eq:optEigenvector}.

Usually, many qubit states or highly entangled states require an ansatz with deep depth. However, the increased dimension of $\Theta_Q$ requires more data to compute the gradient of the cost function. Moreover, a large number of layers can induce a flat landscape of a cost function, which leads to difficulty in training. This is named the Barren plateau~\cite{mcclean2018barren}.
Although various methods are suggested to mitigate the Barren Plateau, being free of this effect seems uneasy. Even with a shallow quantum circuit, QNEE provides useful applications, which will be discussed in Section~\ref{sec:result} in more detail. 

\subsection{Other methods for entropy estimation}
Various VQAs are applicable to estimate quantum entropy~\cite{larose2019variational,tan2021variational, cerezo2022variational}. These algorithms can be classified into two categories. One category is the VQAs based on purity minimization. VQSD is one of these algorithms. VQSD employs the SWAP (swapping two qubits) test to compute the purity~${\rm tr} \{\hat \rho^2 \}$. As the result of minimization, the input density matrix is diagonalized, from which the von Neumann entropy can be estimated. The SWAP test requires $\hat \rho \otimes \hat \rho$ as its input, which means that the width of the quantum circuit must be twice the number of qubits of $\hat \rho$. In addition, another quantum circuit is necessary to compute the cost function.

The other category exploits majorization~\cite{cerezo2022variational}; we say that $\vec x$ majorizes $\vec y$, when $\sum_i^k x_i^{\downarrow} \geq \sum_i^k y_i^\downarrow\quad \text{for arbitrary $k$} $ and $ \sum_i^m x_i =\sum_i^m y_i$ ~\cite{bhatia2013matrix}. Here $m$ is the dimension of the vector, and $\vec x^\downarrow$ is a decreasing ordered vector, i.e., $x^\downarrow_i\geq x^\downarrow_{i+1}$. Because the eigenvalues of $\hat \rho$ majorize the diagonal elements of $\hat \rho$ and the inner product with an ordered energy-level vector is a Schur convex function, the expectation value of an artificially ordered Hamiltonian can be a cost function for diagonalizing $\hat \rho$'s subspace (refer \cite{cerezo2022variational} for detailed information). After optimization, the $\ell$-largest eigenvalues are estimated using the quantum circuit with the optimal parameter~$\Theta^*_Q$. From the string data, we can estimate the eigenvalue $\lambda_i \sim N_i/N_s$ where $N_s$ denotes the number of total shots, and $N_i$ denotes the number of $s_i$ data. %
One benefit of using the cost function is that the requirement for the width of the quantum circuit only needs to match the number of qubits of $\hat \rho$, which is the same requirement as the QNEE. Note that the VQSE focuses on finding a few largest eigenvalues, which makes the VQSE to be well saturated by expanding the allowed solution space of $V(\Theta_Q)$. We note that QNEE's intermediate cost function (see Eq.~\eqref{eq:intermediateCostFtn}) also exploits majorization (see appendix~\ref{sec:Renyiderivation}).
Due to the same structure of the quantum circuit ansatz~$V(\Theta_Q)$ used in QNEE and VQSE, we can fairly compare QNEE with VQSE under the same conditions. We provide this comparison in Sec.~\ref{sec:result}. 

\section{Application of QNEE to a physical model}
\label{sec:result}
\begin{figure*}[th!]
\centering
\includegraphics[width=\textwidth]{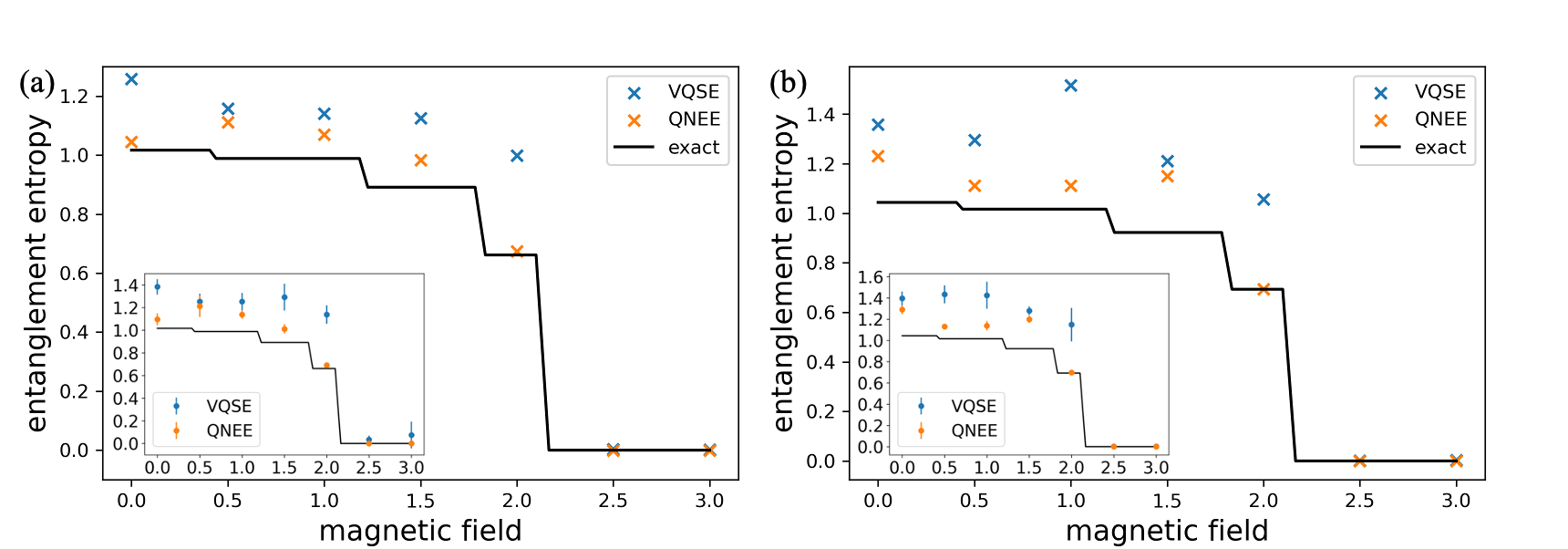}
\caption{ 
Plots of estimation results on the reduced density matrices of the XXZ chain. Total number of qubits is 8. The reduced-density matrix is prepared by partially tracing out (a) five qubits and (b) four qubits. 
The entanglement entropy is estimated with 5 different randomly initialized quantum circuits. `$\times$' symbol represents the lowest value of cost functions among 5 trials. Insets show the mean and standard deviations of 5 trials. Orange-colored symbols represent the results of QNEE. Blue-colored symbols represent the results of VQSE. The exact value of entanglement entropy is denoted by a black line. 
}
\label{fig:magswift}
\end{figure*}

\subsection{Applications of QNEE}
QNEE is a combination of the classical NN and the quantum circuit to estimate the von Neumann entropy. Hence, QNEE shares some advantages of using classical NNs for entropy estimation. For instance, NN such as NEEP has been reported to show high precision in estimating large-valued $f$-divergences, which corresponds to low entropy~(see appendix~\ref{sec:Renyiderivation}), with small-size data or data with rare events~\cite{kim2020learning, SYLee2023}. In this sense, one can expect that QNEE provides high precision for the small-valued von Neumann entropy. While estimating the quantum entropy additionally requires the diagonalization of the density matrix, we note that this does not necessitate a deep depth of the circuit ansatz for low-entropy quantum states.
It is also worth noting that QNEE provides the upper bound Eq.~\eqref{eq:VNcostftn} of the von Neumann entropy even when $\Theta_Q$ is not fully optimized throughout the training process. This guarantees that QNEE efficiently characterizes quantum states with low entropy.

Based on these properties originating from the classical NN, QNEE can be utilized as a phase classifier for quantum states even with a shallow-depth ansatz and a small number of iterations. Particularly, it has been actively studied in many-body physics that phase transitions in various many-body quantum systems accompany the drastic change of the entanglement entropy, given by the von Neumann entropy of the subsystem~\cite{alcaraz1995critical, Osterloh2002Nat, vidal2003entanglement, calabrese2009entanglement, Eisert2010RMP, Bayat2017PRL, Osborne2022PRA}. As an illustrative example, we test our protocol using the XXZ Heisenberg model in a 1D chain (XXZ chain). This model shows the Pokrovsky--Talapov transition that accompanies a sudden change of entanglement entropy~\cite{alcaraz1995critical}. The entanglement entropy abruptly decreases after passing the critical point. 

In the following subsections, we present numerical results of classifying the phase of the XXZ chain using QNEE, and compare its performances with the existing methods of VQSE~\cite{cerezo2022variational}. We employ PyTorch~\cite{paszke2019pytorch}, Qiskit~\cite{mckay2018qiskit}, and QuTiP~\cite{paszke2019pytorch} to implement QNEE.

\subsection{Classifying the phase of XXZ chain}
\label{sec:XXZ_phase}

In order to verify the efficiency of QNEE as a phase classifier, we consider the ground state of the XXZ chain~\cite{park1990finite, alcaraz1995critical, jung2020guide} Hamiltonian,
\begin{align}
    H_{XXZ} = \sum_{l=0}^{L-1} (\hat \sigma_l^x \hat \sigma_{l+1}^x + \hat \sigma_l^y\hat \sigma^y_{l+1} +\Delta \hat \sigma_l^z\hat \sigma^z_{l+1} -\lambda \hat \sigma^z_l ),
\end{align}
where $\hat \sigma_l^k$ denotes $l$-th spin operator in $k \in \{x,y,z\}$ direction and $\Delta$, $\lambda $ are real numbers. $L$ is the total number of qubits in the chain. The last term~$\lambda \hat \sigma^z_l$ can be interpreted as the effects of external magnetic field or chemical potential~\cite{park1990finite}. Here, we impose the periodic boundary condition.

The ground state $\ket{\Psi_G}$ of the XXZ chain shows phase transition at $\lambda_c = 2(1-\Delta )$, which is named Pokrovsky--Talapov tranisiton~\cite{alcaraz1995critical}. For $0 \leq \lambda < \lambda_c$, the system is in the critical or massless phase. In the critical regime, the entanglement entropy shows log scaling behavior~$S \propto \ln{n} + c$, where $n$ is the number of qubits of the reduced matrix, and $c$ is a constant~\cite{vidal2003entanglement}. In the case of the finite-size XXZ chain, the scaling relation is modified as~$S\propto \ln{\{ \frac{L}{\pi}\sin{\frac{\pi n}{L}}  \}} +c'$. For $\lambda > \lambda_c$, the system is in the non-critical or ferromagnetic phase, where the entanglement entropy vanishes. This implies that a sudden decrease of the entanglement entropy can be observed as $\lambda$ increases. 

\begin{figure*}[]
\centering
\includegraphics[width=\textwidth]{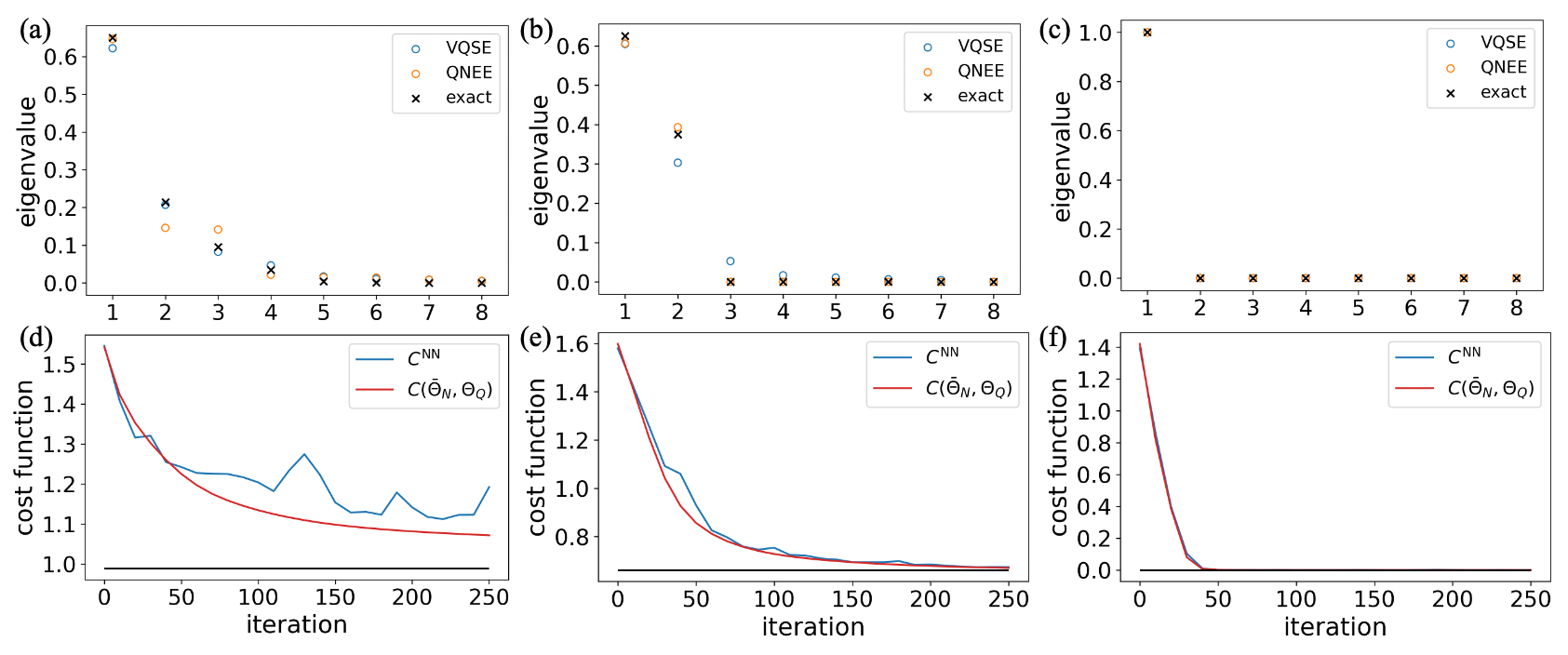} 
\caption{
(a-c) Plots of estimated eigenvalues and (d-f) learning curves for the reduced density matrix with three qubits. These results are from three $\lambda$: (a, d) $\lambda = 0.5$, (b, e) $\lambda = 2.0$, and (c, f) $\lambda = 3.0$. In the top plots, orange (blue) color denotes the estimation results of QNEE (VQSE). $\times$ represents the exact (analytic) result. In the bottom plots, blue line represents cost function~$C^{\rm NN}$ trained with finite samples. The red line represents ideally trained $C(\bar{\Theta}_N,\Theta_Q)$ without finite-sample effect. Here we utilize the same data that is used for Fig~\ref{fig:magswift}(a).
}
\label{fig:eigenvalues}
\end{figure*}

We estimate the entanglement entropy of the XXZ chain with $8$ qubits by estimating the von Neumann entropy of the ground state's reduced density matrix~$\rho = {\rm tr}_B\{ |\Psi_G\rangle \langle \Psi_G|\}$, where ${\rm tr}_B$ denotes partial tracing out the sub system~$B$. We set $\Delta = 0.05$ and vary magnetic field parameter~$\lambda $ from $0$ to $3$. Near the critical point $\lambda \sim 2$, the entanglement entropy undergoes the phase transition. The estimation results of QNEE for the three and four-qubit reduced density matrices are presented in Fig.~\ref{fig:magswift}. We emulate the variational quantum circuit using a classical computer by collecting the outcome strings from the exact probability distribution of the quantum circuit's outcome. The number of layer~$N_l$ is taken as $8$ for the three-qubit case and $10$ for the four-qubit case. The learning rate of gradient-descent method for the quantum circuit is set as $\eta_Q = 0.01$ for both cases. For the NN training process, the learning rate and weight decay of ADAM optimizer are set as $0.00001$ and $0.00005$, respectively. We train the NN over $N_{\rm iter} = 10^4$ iterations at an initial $\Theta_Q$ for sufficient optimization. 
In the subsequent NN training processes, the NN is trained over $N_{\rm iter} = 100$ iterations at each $\Theta_Q$. The entanglement entropy is estimated from five different initial $\Theta_Q$, and the minimum value among them is taken. In the insets of Fig.~\ref{fig:magswift}, we plot the mean and standard deviation of the five different trials. For each $\Theta_Q$, 30,000 shots are used to find the optimal parameter $\bar \Theta_N$.

The numerical results demonstrate that QNEE well estimates the entanglement entropy of the XXZ chain ground state for a wide range of the parameter regime. In particular, QNEE provides an accurate estimation near $\lambda_c$ and the non-critical regime. This is expected as these are low-entropy states that a classical NN can estimate well. In particular, the standard deviation of QNEE is smaller than the size of markers when $\lambda \geq \lambda_c$, which means that a single trial is enough to classify those phases with QNEE. 

Another output of QNEE is the eigenvalues of the reduced density matrix, the so-called entanglement spectrum. In Fig.~\ref{fig:eigenvalues}~(a - c), we plot the estimated eigenvalues for three different values of magnetic field~$\lambda =0.5, 2.0, 3.0 $ for the three-qubit reduced density matrix using the same data as in Fig.~\ref{fig:magswift}. Even though QNEE mainly focuses on entropy estimation, its eigenvalue estimation is comparable to VQSE, which is designed for this particular task. A more detailed comparison between QNEE and VQSE will be discussed in the following subsection.

In Fig.~\ref{fig:eigenvalues}~(d-f), we plot the learning curves of QNEE for the case of three qubits. The blue line represents the learning curve with the cost function calculated by the classical NN using a finite number of samples. The red curve represents the learning curve with the cost function obtained from the exact wave function of the ground state. As shown in Fig.~\ref{fig:eigenvalues}, $100\sim200$ iterations are enough to classify the phases. A similar number of iterations are also enough for the four-qubit case.

\subsection{Comparison between QNEE and VQSE}
\label{sec:compare_QNEE_VQSE}

We compare the performance of QNEE with VQSE in more detail. We first clarify that the cost functions of QNEE and VQSE are designed for different purposes; the former is for entropy estimation, while the latter is for estimating $\ell + 1$ lowest eigenvalues (see section `Details of VQSE' of the ``Results" and Ref.~\cite{cerezo2022variational} for more details). Therefore, it is expected that QNEE will provide better performance in the entropy estimation task, as observed in Fig.~\ref{fig:magswift}. On the other hand, the eigenvalue estimation by VQSE is, in general, more accurate than QNEE.

In our XXZ chain model, both QNEE and VQSE show good agreement in entropy/eigenvalue estimation in the non-critical regime ($\lambda > \lambda_c$), where the entropy is vanishing. Near the transition point ($\lambda \sim \lambda_c$), QNEE produces a more reliable entropy estimation than VQSE, while the accuracies of the eigenvalue estimations are comparable. In the critical regime, ($\lambda < \lambda_c$) QNEE estimates the entropy more accurately than VQSE. In particular, for $\lambda = 0.5$, VQSE estimates the eigenvalues (see Fig.~\ref{fig:eigenvalues}(a)) better than QNEE, but QNEE yields a higher accuracy of the estimated entropy than VQSE (see Fig.~\ref{fig:magswift}(a)). This could lead to an intriguing implication that, in certain cases, the von Neumann entropy can be accurately estimated using QNEE, even when the estimated eigenvalues do not exactly match the correct values. On the other hand, the main advantage of VQSE comes from its low computational cost. As VQSE does not involve a training process, it runs much faster than QNEE to estimate the quantum state's entropy and eigenvalues while yielding a comparable performance.

\begin{figure}[t]
\centering
\includegraphics[width=0.5\textwidth]{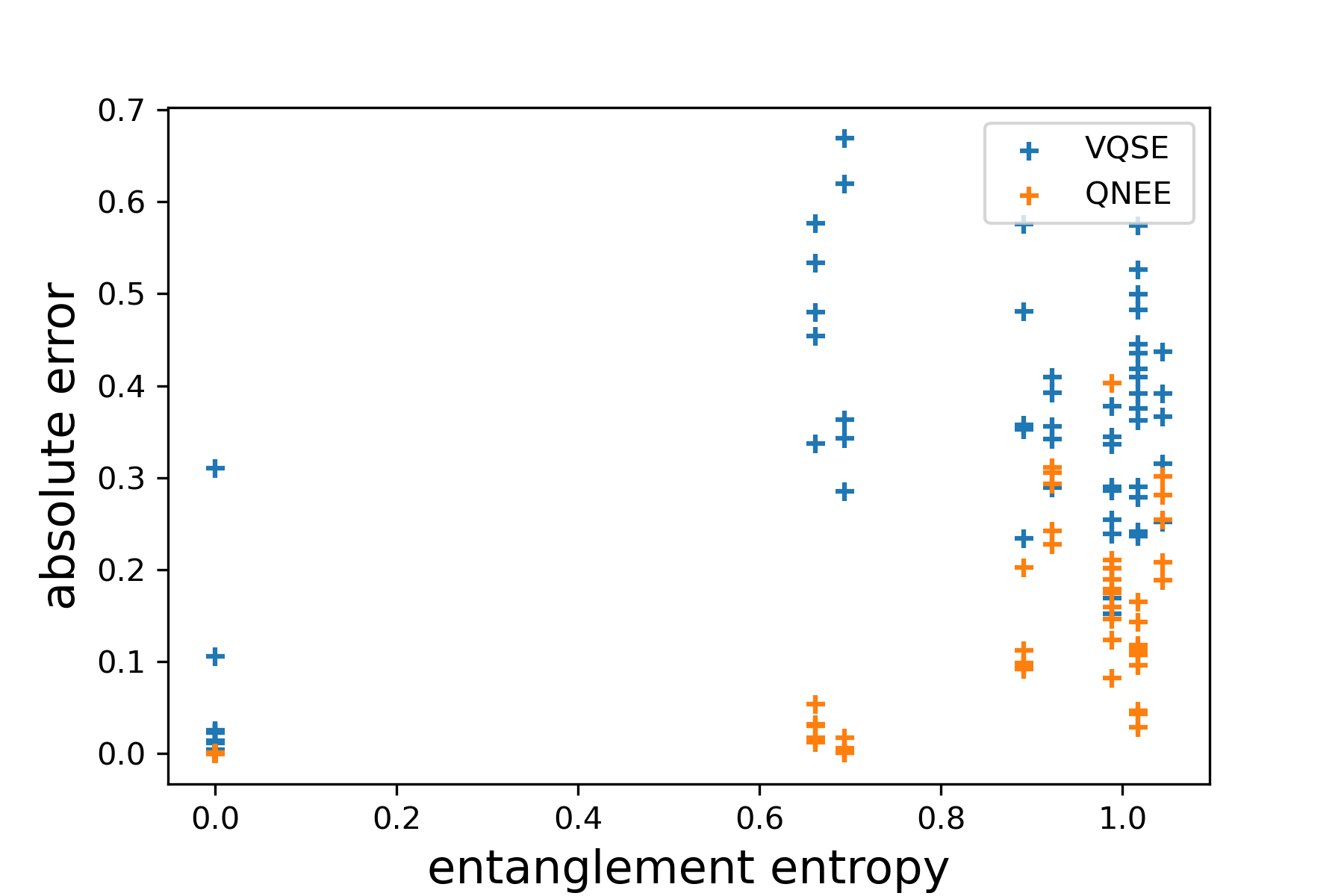}
\caption{ 
Scatter plot of the absolute errors of QNEE (orange symbol) and VQSE (blue symbol) as a function of entanglement entropy. For this plot, we take data from reduced density matrices of three and four qubits of the XXZ chain. 
}
\label{fig:errorplot}
\end{figure}

We also note that QNEE is less sensitive to the initial choices of $\Theta_Q$ than VQSE. This can be captured by comparing the variance of estimated entropy by taking different initial points of $\Theta_Q$ (see the insets of Fig.~\ref{fig:magswift}). The estimator's variance for QNEE ranges up to $0.09$. In contrast, the variance of VQSE ranges up to $0.16$, which is significantly higher than that of QNEE. From the simulation data, we further analyze the behavior of errors depending on the exact entropy value. In Fig.~\ref{fig:errorplot}, we plot the absolute error of estimation results of QNEE~$|S(\hat \rho) - C(\Theta_N, \Theta_Q)|$ as a function of entanglement entropy for $3$- and $4$-qubit reduced density matrices of the XXZ chain. QNEE results show a clear tendency where the error and its fluctuation decrease as the entanglement entropy decreases (see Fig.~\ref{fig:errorplot}). While VQSE shares a similar tendency, this is not as clear as the QNEE case.

To summarize, QNEE performs better in entropy estimation, while VQSE is more specialized in finding the first few largest eigenvalues. We expect that such a comparison will also be valid for general cases with a larger number of qubits, where a considerable number of non-zero eigenvalues are involved in the quantum state's entropy. However, we must note that the observed tendency between the two methods is not definite and may depend on the quantum state of interest and the choice of variational parameters in quantum circuit ansatz and classical optimizers.

\section{Discussion}
\label{sec:disscussion}
We have proposed a novel quantum entropy estimator, QNEE, which combines variational quantum circuits with classical NNs. Our method allows us to estimate the von Neumann entropy and the R\'enyi entropy from finite samples of circuit outcomes without accessing full information of the density matrix. By leveraging the Donsker-Varadhan inequality to optimize the cost function, classical NNs in the QNEE inherit valuable characteristics from recently developed classical entropy estimators, including NEEP~\cite{kim2020learning}. We emphasize that classical NNs operate independently of the structure of variational quantum circuits, offering a universal application to any form of quantum circuit ansatz.

As a physical model, we have applied QNEE to estimate the entanglement entropy of the ground state. We have demonstrated that QNEE effectively classifies the massless and ferromagnetic phases of the 1D XXZ Heisenberg model. In particular, numerical evidence shows that QNEE can sensitively estimate the entanglement entropy near the critical point even before the estimated eigenvalues converge to the exact ones. We also highlight that such phase classification can still work even without accurately estimating entanglement entropy when one of the phases has low entanglement, as the estimator is designed to upper bound the exact entropy value with sufficient samples. 

There is a caution when applying QNEE. The performance of QNEE is sensitive to the learning rate of the gradient-descent method for the quantum circuit. A lower learning rate is recommended as the number of qubits increases or the number of shots decreases.
This can be a potential issue when applying the QNEE to a high dimensional quantum state, as a high learning rate can induce fluctuations in the learning curve.
%

%

Our methodology establishes a pathway for investigating quantum state properties using fewer iterations of quantum circuits, which are considerably more resource-intensive than classical processing. We anticipate that QNEE will serve as a vital tool for addressing a broad spectrum of estimation and classification challenges within quantum systems.

\begin{acknowledgments}
The authors acknowledge the Korea Institute for Advanced Study for providing computing resources (KIAS Center for Advanced Computation Linux Cluster System). This research was supported by the KIAS Individual Grant Nos. PG081801 (S.L.), CG085301 (H.K.), and PG064901 (J.S.L.) at the Korea Institute for Advanced Study. S. L. thanks Dong Kyum Kim for helpful discussions about machine learning.
\end{acknowledgments}

\appendix
\renewcommand\thefigure{\thesection.\arabic{figure}}
\setcounter{figure}{0}

\section{Derivation of the upper bound of the von Neumann entropy}
\label{sec:derivation}
Here, we derive the upper bound of the von Neumann entropy, i.e., Eq.~\eqref{eq:costfunction}.
In terms of the eigenstate~$|\lambda_m\rangle$ of a density matrix $\hat \rho$, that is $\hat \rho = \sum_m \lambda_{m} |\lambda_{m}\rangle \langle \lambda_{m}|$, one can write an inequality for an observable $\hat O$ as follows: 
\begin{align}
    \ln{\tr{ e^{\hat O} }} =& \ln{
    \sum_m  \langle \lambda_{m} |e^{\hat O} |\lambda_{m}\rangle
    }\nonumber\\
    \geq& \ln{
    \sum_m  e^{\langle \lambda_{m} |\hat O| \lambda_{m} \rangle} } \nonumber \\
    = &\ln{
    \sum_m \lambda_m e^{\langle \lambda_{m} |\hat O| \lambda_{m} \rangle - \ln{\lambda_m} } } \nonumber \\
    \geq&\ln{
    e^{ \sum_m \lambda_m \left( \langle \lambda_{m} |\hat O| \lambda_{m} \rangle - \ln{\lambda_m} \right) } }\nonumber\\ 
    =& \langle \hat O\rangle_{\hat{ \rho}  } + S(\hat \rho). \label{eqA:inequality_derivation}
\end{align}
Convexity of the exponential function is used for deriving the inequalities in the second and the fourth lines of Eq.~\eqref{eqA:inequality_derivation}. We note that the second-line inequality is saturated when the observable operator is diagonalized by the basis set $\{|\lambda_{m}\rangle \}$. Equation~\eqref{eqA:inequality_derivation} can be rearranged as 
\begin{align}
    S(\hat \rho )\leq -\langle \hat O\rangle_{\hat \rho} +\ln{ {\rm tr} \{e^{\hat O}\}  },
\label{eq:GibbsVar}\end{align} 
which represents the Gibbs variational principle. In general, the Gibbs variational principle focuses on finding an optimal density matrix, enabling Eq.~\eqref{eq:GibbsVar} to be saturated, which is given by the
Gibbs state~$\hat \rho^* = e^{\hat O}/ {\rm tr} \{ e^{\hat O}\} $. In contrast, QNEE focuses on finding an optimal observable achieving the equality of Eq.~\eqref{eq:GibbsVar}, which is given by~$\hat O^* = \ln{ \hat \rho }$. 

For implementing the variational method to obtain the optimal entropy value via NN, one might be able to use the right-hand side (rhs) of Eq.~\eqref{eq:GibbsVar} as a cost function. However, when employing the stochastic (mini-batch) gradient descent method for optimizing the parameters of the NN, the cost function, including a logarithmic function such as the rhs of Eq.~\eqref{eq:GibbsVar} can malfunction and yield an incorrect result without additional data manipulation~\cite{belghazi2018mine}.
To avoid this complication, we linearize the logarithmic function by applying the additional inequality, ${\rm tr} \{e^{\hat O}\}-1\geq \ln {\rm tr} \{e^{\hat O}\}$, which is also saturated when $\hat O^* = \ln \hat \rho$. Application of this additional inequality to Eq.~\eqref{eq:GibbsVar} leads to Eq.~\eqref{eq:costfunction} as follows:  
\begin{align}
    S(\hat \rho )\leq -\langle \hat O\rangle_{\hat \rho} +{\rm tr}{ \{e^{\hat O}\}  } -1.
\label{eq:finVNform}
\end{align} 

\section{Derivation of the upper bound of the R\'enyi entropy}
\label{sec:Renyiderivation}
Here, we derive the upper bound of the R\'enyi entropy in Eq.~\eqref{eq:Renyicost_maintext}. For a convex and twice-differentiable real-valued function $f(x)$ with arbitrary real numbers $p$ and $q ~(\neq 0)$, the following inequality
\begin{align} \label{eqB:baseinequality}
    -q f(p/q) \leq -pf'(x) + q \left\{ xf'(x) - f(x) \right\} 
\end{align}  
holds~\cite{Kwon2023Divergence}. We can demonstrate this by differentiating the rhs of Eq.~\eqref{eqB:baseinequality} with respect to $x$, which yields $f''(x)(-p+q x)$. As $f''(x) >0$ by convexity, the rhs has the unique minimum $-q f(p/q)$ at $x= p/ q$. As Eq.~\eqref{eqB:baseinequality} is valid for arbitrary $p$ and $q$, we can replace $p$ and $q$ with some distributions $P(s_i)$ and $Q(s_i)$, where $\sum_i P(s_i) = \sum_i Q(s_i) = 1$. By considering that $x=x(s_i;\theta)$ is a function of $s_i$ and parameterized by $\theta$, and summing over all $i$, we can write Eq.~\eqref{eqB:baseinequality} as the following variational form: 
\begin{widetext}
    
\begin{align}
    - D_{f}(P \| Q) \leq \sum_{i} 
    \left[
        -P(s_i) f'\left( x(s_i ;\theta) \right) + Q(s_i) \left\{x(s_i ;\theta) f'( x(s_i;\theta) ) - f( x(s_i;\theta) )  \right\}
    \right]  ,
\label{eq:ineqdiscrete}\end{align}
\end{widetext}
where $D_{f}(P \| Q) \equiv \sum_{i} Q(s_i)f\left( P(s_i) / Q(s_i) \right)$ is  the $f$-divergence.
To connect this inequality to the R\'enyi entropy, we choose the convex function $f(x)$ as
\begin{align} \label{eqB:f_function}
    f(x) = \frac{x^\alpha - \alpha x -(1-\alpha) }{\alpha(\alpha -1)}
\end{align} 
for $\alpha \neq 0,1$. Then, the $f$-divergence can be rewritten as 
\begin{align}
    D_f(P\| Q) = \frac{1}{\alpha(\alpha - 1 )}
    \left [\sum_{i} \frac{P^\alpha(s_i)}{Q^{\alpha-1}(s_i)} - 1\right ]. 
\label{eq:fdiv2}\end{align}
We can make a relation between the $f$-divergence and the R\'enyi-divergence $D_\alpha (P\| Q)$ as
\begin{align} \label{eqB:f_and_Renyi_div}
    D_\alpha (P\| Q) \equiv& \frac{1}{\alpha -1} \ln{\left[ \sum_{i}\, \frac{P^\alpha(s_i)}{Q^{\alpha-1}(s_i)} \right]}\nonumber\\ 
    =& \frac{1}{\alpha -1}  \ln{\left[ \alpha (\alpha -1) D_f(P \| Q) + 1 \right]}.
\end{align}
By choosing $Q$ as the uniform distribution, i.e., $Q(s_i)= 1/d$ for all $i$ with $d = \sum_i 1$, the R\'enyi divergence is reduced to the R\'enyi entropy as  
\begin{align} \label{eqB:Renyi_div_and_ent}
    D_\alpha (P\| Q) = - H_\alpha(P) +\ln{d},
\end{align}
where the R\'enyi entropy is defined as $H_\alpha (P) \equiv \frac{1}{1-\alpha} \ln{\sum_{i} P^\alpha(s_i) }$. Plugging Eqs.~\eqref{eqB:f_function}, \eqref{eqB:f_and_Renyi_div}, and \eqref{eqB:Renyi_div_and_ent} into Eq.~\eqref{eq:ineqdiscrete}, and substituting $x(s_i;\theta)$ with $N e^{h(s_i;\theta)}$, Eq.~\eqref{eq:ineqdiscrete} can be arranged as 
\begin{align}
    \frac{e^{ (1-\alpha) H_\alpha(P)  } }{\alpha (1-\alpha)} \leq \sum_{i}
       P(s_i)\frac{ e^{(\alpha -1)h(s_i;\theta) } }{1 - \alpha }
    +\sum_{i}
        \frac{e^{\alpha h(s_i;\theta)} }{\alpha} .
\label{eq:CRenyicost}\end{align}
The quantum generalization of the R\'enyi entropy for a given density matrix $\hat\rho$ is defined as $S_\alpha(\hat\rho) \equiv \frac{1}{1-\alpha} \ln \tr{\hat\rho^\alpha} = \frac{1}{1-\alpha} \ln \left( \sum_i \lambda_i^\alpha \right)$. By noting that the probability distribution $P_i = \tr{\hat \rho \Pi_i}$ for a set of measurement operators $\{ \Pi_i \}$ is always majorized by the distribution of eigenvalues $\lambda_i$ and that the R\'enyi entropy is a Shcur-concave function, we obtain
$$
H_\alpha(P) \geq S_\alpha(\hat\rho).
$$
In particular, the inequality is saturated when the measurement basis is the eigenbasis of the density matrix $\hat\rho$. By taking $P_{\hat V}(s_i;\Theta_Q) = \langle s_i | \hat V(\Theta_Q) \hat \rho \hat V^\dagger (\Theta_Q) |s_i \rangle  = \tr{ \hat\rho_{\hat V} \Pi_i}$ and subtracting $1/[\alpha(1-\alpha)]$ from both sides of Eq.~\eqref{eq:CRenyicost}, we can reach the final result as
\begin{align}
    \frac{e^{ (1-\alpha) S_\alpha( \hat \rho )  } -1 }{\alpha (1-\alpha)} 
     \leq&
     \frac{e^{ (1-\alpha) H_\alpha( P_{\hat V}(s_i; \Theta_Q))  } -1 }{\alpha (1-\alpha)} 
    \nonumber\\ 
    \leq&
    \sum_{i}
    P_{\hat V}(s_i;\Theta_Q) \frac{ e^{(\alpha -1)h(s_i;\Theta_N) } -1}{(1 - \alpha )}\nonumber\\
    &+\sum_{i} \frac{
        e^{\alpha h(s_i;\Theta_N)} - 1}{\alpha},
    \label{eq:Renyicost}
\end{align}
which is Eq.~\eqref{eq:Renyicost_maintext}. Note that Eq.~\eqref{eq:Renyicost} is reduced to the inequality for the von Neumann entropy~\eqref{eq:VNcostftn} in the $\alpha \rightarrow 1$ limit. 

\section{details of VQSE}
\label{sec:VQSE}
In this section, we illustrate the details of VQSE~\cite{cerezo2022variational}. VQSE exploits the relationship between majorization and diagonalization. The cost function is given as the expectation value of an artificial Hamiltonian
\begin{align}
   \hat H \equiv& (1-t)\hat H_L + t\hat H_G, 
\end{align}
where $ \hat H_L \equiv \hat I - \frac{1}{2}\sum^\ell_{j=1} r_j \hat \sigma_j^z $ is local Hamiltonian that is designed to mitigate Barren plateau, and $\hat H_G \equiv \hat I - \sum_{s_j\in S} q_j |s_j\rangle \langle s_j|$ is a global Hamiltonian. $t$, $r_j$, and $q_j$ are real number parameters. $S$ is the set of $m$ most numerous strings from the quantum circuit, where $m$ is a positive integer less than $2^\ell$. At the start of the optimization process, $t=0$ and $t$ increases at every certain iteration up to $t = 1$. The set~$S$ is also updated with the same period with $t$. $\hat I$ is the identity operator, and $\ell$ is the number of qubits. There exist many choices for $q_j$ and $r_j$. The choice used in reference~\cite{cerezo2022variational} is $r_j = r_1 + (j-1)\delta $, and the eigenvalues of $\hat H_G$ are set to be the same value as $\ell + 1$ lowest eigenvalues of $\hat H_L$. The other eigenvalues of $\hat H_G$ are set to be $1$. This choice of $r_j$ ensures $\ell+1$ lowest non-degenerate eigenvalues of $\hat H_L$. Therefore, $\ell +1 $ eigenvalues can be estimated. In our research, we set $r_1= 0.2$, $\delta=0.01$, and $\ell =n$. We set the learning rate for training VQSE to be $0.05$ and update $t$ and $S$ at every $25$ iterations.

\bibliography{entangle}

\end{document}